\documentclass[11pt,epsf]{article}
\usepackage{amsmath}
\usepackage{amsfonts}
\usepackage{amssymb}
\usepackage{graphicx}
\newtheorem{theorem}{Theorem}

\newcommand {\hx} {\hat{x}}
\newcommand {\hX} {\hat{X}}
\newcommand {\dfn} {\stackrel{\Delta} {=}}

\newcommand {\reals} {{\rm I\!R}}

\newcommand{\calA}{{\cal A}}

\newcommand{\calE}{{\cal E}}
\newcommand{\calF}{{\cal F}}
\newcommand{\calG}{{\cal G}}
\newcommand{\calH}{{\cal H}}

\newcommand{\calP}{{\cal P}}

\newcommand{\calS}{{\cal S}}

\newcommand{\calU}{{\cal U}}
\newcommand{\calV}{{\cal V}}
\newcommand{\calW}{{\cal W}}
\newcommand{\calX}{{\cal X}}
\newcommand{\calY}{{\cal Y}}

\begin{document}
\thispagestyle{empty}
\title{An Identity of Chernoff Bounds with an Interpretation
in Statistical Physics and Applications in Information Theory
}
\author{Neri Merhav
}
\maketitle

\begin{center}
Department of Electrical Engineering \\
Technion - Israel Institute of Technology \\
Haifa 32000, Israel \\
\end{center}
\vspace{1.5\baselineskip}
\setlength{\baselineskip}{1.5\baselineskip}

\begin{abstract}

An identity between two versions of the 
Chernoff bound on the probability a certain
large deviations event, is established. This identity has an interpretation
in statistical physics, namely, an isothermal equilibrium of a composite system that
consists of multiple subsystems of particles. 
Several information--theoretic application examples, where
the analysis of this large deviations probability naturally arises, are 
then described from the viewpoint of this statistical mechanical interpretation.
This results in several relationships between 
information theory and statistical physics, which
we hope, the reader will find insightful.

\vspace{0.5cm}

\end{abstract}

\section{Introduction}

Relationships between information theory and statistical physics have been
extensively recognized over the last few decades, and they are drawn from
many different aspects. We mention here only a few of them.

One such aspect is characterized by identifying structures
of optimization problems pertaining to certain information--theoretic settings
as being analogous to parallel structures that arise in statistical physics,
and then borrowing statistical--mechanical 
insights, as well as powerful analysis techniques (like the replica
method) from statistical physics to the dual information--theoretic setting
of interest. A very partial list of works along this line includes 
\cite{AB01}, 
\cite{GV02},
\cite{HK05},
\cite{KH05},
\cite{KNM02},
\cite{KabS99}
\cite{KSNS01},
\cite{KanS99},
\cite{MM06} (and references therein),
\cite{MR06},
\cite{Murayama02},
\cite{PS99},
\cite{RC00},
\cite{Sourlas89},
\cite{Sourlas94},
\cite{Tanaka01},
\cite{Tanaka02},
and \cite{WSW05}.

Another aspect pertains to the 
philosophy and the application of the maximum entropy principle,
which emerged in statistical mechanics 
in the nineteenth century and has been advocated during the
previous century in
a wide variety of more general contexts, by Jaynes
\cite{Jaynes57a},\cite{Jaynes57b},\cite{Jaynes82}, and by 
Shore and Johnson \cite{SJ80}, as a general guiding principle
to problems in information theory 
(see, e.g., \cite[Chap.\ 11]{CT91} and references therein) 
and other areas, such as signal processing, 
in particular, speech coding (see, e.g., \cite{GGRS81})
spectrum estimation (see, e.g., \cite{Burg75}), and others.

Yet another aspect is related to ideas and theories that 
underly the notion of `trading' between
information bits and energy, or heat. In particular, 
Landauer's erasure principle 
\cite{Landauer61} is argued to provide a powerful link between  
information theory and physics and to 
suggest a physical theory of information
(comprehensive overviews are included in, 
e.g., \cite{Maroney04} and \cite{PV01}). 
According to Landauer's principle, the erasure of
every bit of information increases the thermodynamic 
entropy of the world by $k\ln 2$, where $k$ is Boltzmann's
constant, and so, information is actually physical. 

Finally, to shift gears more to the direction of this paper,
we should mention the aspect of the interface between statistical physics and 
large deviations theory,
a line of research advocated most 
prominently by Ellis \cite{Ellis85},\cite{Ellis06},
and developed also by Oono \cite{Oono89}, 
McAllester \cite{McAllester}, and others. The main
theme here evolves around the 
identification of Chernoff bounds and more general large deviations
rate functions with free energies (along with
their related partition functions),
thermodynamical entropies, and the underlying maximum--entropy/equilibrium principle associated with them.
In particular, Ellis' book \cite{Ellis85}
is devoted largely to the application of large deviations theory 
to the statistical physics pertaining to 
models of ferromagnetic spin arrays, like 
Ising spin glasses and others, 
in order to explore
phase transitions phenomena of spontaneous 
magnetization (see also \cite{MM06}).

This paper, which is mostly expository in character,
lies in the intersection 
of information theory, large deviations theory, and
statistical physics. In particular, we establish a simple identity between two
quantities as they can both be interpreted as the rate
function of a certain large deviations event
that involves multiple 
distributions of sets of independent random variables (as opposed
to the usual, single set of i.i.d.\ random variables).
The analysis of this large deviations event is of a general form
that is frequently encountered in numerous applications in
information theory (cf.\ Section 4). Its informal description is as follows:
Let $v_1,\ldots,v_n$ be an arbitrary 
(deterministic) sequence whose components take
on values in a finite set $\calV$, and let $U_1,\ldots,U_n$ be a sequence of
random variables where each component is generated independently
according to a distribution $q(u_i|v_i)$, $i=1,\ldots,n$.
For a given function $f$ and a constant $E$,
we are interested in the large deviations
analysis (Chernoff bound) of the probability of the event
\begin{equation}
\label{event}
\sum_{i=1}^n f(U_i,v_i)\le nE,
\end{equation}
assuming that the relative frequencies of the various symbols in
$(v_1,\ldots,v_n)$ stabilize as $n$ grows without bound, and assuming
that $E$ is sufficiently small to make this a rare event for large $n$.

There are (at least) two ways to drive a Chernoff bound on the probability
of this event. The first is to treat the entire sequence of RV's,
$\{f(U_i,v_i)\}$ as a whole, and the second is to partition it
according to the various symbols $\{v_i\}$, i.e., to consider the separate
large deviations events  of the 
partial sums, $\sum_{i:v_i=v}f(U_i,v)$, $v\in\calV$, for all possible
allocations of the total `budget' $nE$ among the various $\{v\}$.
These two approaches lead to two 
(seemingly) different expressions of Chernoff bounds,
but since they are both exponentially tight, they must agree.

As will be described and discussed in Section 2, 
the identity between these two Chernoff bounds has a natural
interpretation in statistical physics: it is viewed as 
a situation of thermal equilibrium (maximum
entropy) in a system that consists of several 
subsystems (which can be of different kinds), each of them with many particles.

As will be shown in Section 4,
the above--described problem of large deviations analysis of the
event (\ref{event}) is encountered in many applications in information
theory, such as rate--distortion coding, channel capacity, hypothesis
testing (signal detection, in particular), and others. The
above mentioned statistical mechanical 
interpretation then applies to all
of them. Accordingly, Section 4 is devoted to expository descriptions of
each of these applications, along with 
the underlying physics that is inspired by
the proposed thermal equilibrium interpretation. The reader is assumed to have
very elementary background in statistical physics.

The remaining part of this paper is organized as follows. In Section 2,
we establish some notation conventions. In Section 3, we assert and prove 
our main result, which is the identity between the above described
Chernoff bounds. Finally, in Section 4, we explore the application
examples.

\section{Notation}

Throughout this paper, scalar random 
variables (RV's) will be denoted by the capital
letters, like $U$,$V$,$X$, and $Y$, their sample values will be denoted by
the respective lower case letters, and their alphabets will be denoted
by the respective calligraphic letters.
A similar convention will apply to
random vectors and their sample values,
which will be denoted with same symbols superscripted by the dimension.
Thus, for example, $X^n$ will denote a random $n$-vector $(X_1,\ldots,X_n)$,
and $x^n=(x_1,...,x_n)$ is a specific vector value in $\calX^n$,
the $n$-th Cartesian power of $\calX$. The 
notations $x_i^j$ and $X_i^j$, where $i$
and $j$ are integers and $i\le j$, will designate segments $(x_i,\ldots,x_j)$
and $(X_i,\ldots,X_j)$, respectively,
where for $i=1$, the subscript will be omitted (as above).
Sequences without specifying indices are denoted by $ \{\cdot\} $.
Sources and channels will be denoted generically by the letter $P$ or $Q$.
Specific letter probabilities corresponding to a source $P$ will be
denoted by the corresponding lower case letter, e.g., $p(v)$ is the
probability of a letter $v\in\calV$. A similar convention will be applied
to a channel $Q$ and the corresponding transition probabilities, e.g., $q(u|v)$,
$u\in\calU$, $v\in\calV$.
The cardinality of a finite set $\calA$ will be denoted by $|\calA|$.
Information theoretic quantities like entropies, and mutual
informations will be denoted following the usual conventions
of the information theory literature.

Notation pertaining to statistical physics 
will also follow, wherever possible,
the customary conventions. I.e., $k$ will denote 
Boltzmann's constant ($k=1.38065\times 10^{-23}$ Joules
per Kelvin degree), $T$ -- absolute temperature (in Kelvin 
degrees), $\beta=1/(kT)$ -- the inverse temperature
(in units of $\mbox{Joule}^{-1}$ or $\mbox{erg}^{-1}$), 
$E$ -- energy, the letter $Z$ will be used to
denote partition functions, etc.

\section{Main Result}

Let $\calU$ and $\calV$ be finite\footnote{The assumption that $\calU$ is
finite, is made mostly for the sake of convenience and simplicity. Most
of our results extend straightforwardly to the case of a continuous
alphabet $\calU$. The extension to a continuous alphabet $\calV$ is somewhat
more subtle, however.}
sets and let $f:\calU\times\calV\to\reals$
be a given function. Let 
$P=\{p(v),~v\in\calV\}$ 
be a probability mass function on $\calV$ and 
let $Q=\{q(u|v),~u\in\calU,~v\in\calV\}$ be a 
matrix of conditional probabilities from $\calV$ to $\calU$.

Next, let us define for each $v\in\calV$,
the partition function:
\begin{equation}
\label{zvb}
Z_v(\beta)=\sum_{u\in\calU}q(u|v)e^{-\beta f(u,v)},~~~~\beta\ge 0,
\end{equation}
and for a given $E_v$ in the range
\begin{equation}
\label{range}
\min_{u\in\calU}f(u,v) \le E_v \le
\sum_{u\in\calU}q(u|v)f(u,v),
\end{equation}
let
\begin{equation}
S_v(E_v)=\min_{\beta\ge 0}[\beta E_v+\ln Z_v(\beta)].
\end{equation}
Further, for a given constant $E$ in the range
$$\sum_{v\in\calV}p(v)\min_{u\in\calU}f(u,v) \le E \le
\sum_{u\in\calU}\sum_{v\in\calV}p(v)q(u|v)f(u,v),$$ 
let
\begin{equation}
\bar{S}(E)=\min_{\beta\ge 0}\left[\beta E+
\sum_{v\in\calV}p(v)\ln Z_v(\beta)\right].
\end{equation}
Let $\calH(E)$ denote the set of all $|\calV|$--dimensional vectors 
$\bar{E}=\{E_v,~v\in\calV\}$, where each component $E_v$ satisfies 
(\ref{range}),
and where $\sum_vp(v)E_v\le E$.
Our main result, in this section, is the following:

\begin{theorem}
\begin{equation}
\label{identity}
\max_{\bar{E}\in\calH(E)}\sum_{v\in\calV}p(v)S_v(E_v)=\bar{S}(E).
\end{equation}
\end{theorem}

The expression on the right--hand side is, 
of course, more convenient to work with since
it involves minimization w.r.t.\ one parameter 
only, as opposed to the left--hand side,
where there is a minimization over $\beta$ 
for every $v$, as well as a maximization
over the $|\calV|$--dimensional vector $\bar{E}$.

While the proof of Theorem 1 below is fairly short, 
in the Appendix (subsection A.1), we outline an alternative
proof which, although somewhat longer, 
provides some additional insight, we believe.
As described briefly in the Introduction, 
it is based on two different approaches to the analysis 
of the rate function, $I(E)$, pertaining to
the probability of the event:
\begin{equation}
\label{ld}
\sum_{i=1}^n f(U_i,v_i)\le nE,
\end{equation}
where $\{U_i\}$ are RV's taking values in $\calU$ and drawn according to 
$q(u^n|v^n)=\prod_{i=1}^nq(u_i|v_i)$, and
$v^n=(v_1,\dots,v_n)$ is a given deterministic vector whose 
components are in $\calV$,
with each $v\in\calV$ appearing
$n_v$ times ($\sum_{v\in\calV}n_v=n$), and
the related relative frequency, $n_v/n$ is exactly $p(v)$.

It should be noted that 
the proof in the Appendix pertains to a 
slightly different definition of the set
$\calH(E)$, where the individual upper bound to 
each $E_v$ is enlarged to $\max_uf(u,v)$.
Thus, $\calH(E)$ is extended to a larger set, 
which will be denoted by $\calH_0(E)$ in the Appendix. But the
maximum over $\calH_0(E)$ is 
always attained within the original set $\calH(E)$
(as is actually shown in the proof below).

\vspace{0.5cm}

\noindent
{\it Proof.}
Here we prove the identity of Theorem 1 
directly, without using large deviations analysis and Chernoff bounds.
We first prove that for every $\bar{E}\in\calH(E)$, 
we have $\sum_{v\in\calV}p(v)S_v(E_v)\le \bar{S}(E)$
and then, of course,
$$\max_{\bar{E}\in\calH(E)}\sum_{v\in\calV}p(v)S_v(E_v)\le \bar{S}(E)$$ 
as well. This follows from the following chain of inequalities:
\begin{eqnarray}
\sum_{v\in\calV}p(v)S_v(E_v)&=&\sum_{v\in\calV}p(v)\cdot\min_{\beta\ge 0}[\beta E_v+\ln Z_v(\beta)]\nonumber\\
&=&\sum_{v\in\calV}\min_{\beta\ge 0}[\beta p(v)E_v+p(v)\ln Z_v(\beta)]\nonumber\\
&\le&\min_{\beta\ge 0}\left[\beta\sum_{v\in\calV}p(v)E_v+
\sum_{v\in\calV}p(v)\ln Z_v(\beta)\right]\nonumber\\
&\le&\min_{\beta\ge 0}\left[\beta E+\sum_{v\in\calV}p(v)\ln Z_v(\beta)\right]\nonumber\\
&=&\bar{S}(E),
\end{eqnarray}
where in the second inequality we used the postulate that
$\sum_vp(v)E_v\le E$.

In the other direction, let $\beta^*$ be the achiever of $\bar{S}(E)$, 
i.e., $\beta^*$ is the solution to the equation:
$$E=-\left[\frac{\partial}{\partial\beta}\sum_vp(v)
\ln Z_v(\beta)\right]_{\beta=\beta^*}.$$
For each $v\in\calV$, 
let $E_v^*\in[\min_uf(u,v),\sum_uq(u|v)f(u,v)]$ be chosen such that
$\beta^*$ would be the achiever of $S_v(E_v^*)$, i.e., $E_v^*=-[\partial\ln Z_v(\beta)/\partial\beta]_{\beta=\beta^*}$.
Obviously, the vector $\{E_v^*,~v\in\calV\}$ lies in $\calH(E)$, and
\begin{eqnarray}
\sum_vp(v)E_v^*&=&-\sum_vp(v)\left[\frac{\partial\ln 
Z_v(\beta)}{\partial\beta}\right]_{\beta=\beta^*}\nonumber\\
&=&-\left[\frac{\partial}{\partial\beta}\sum_vp(v)
\ln Z_v(\beta)\right]_{\beta=\beta^*}\nonumber\\
&=&E.
\end{eqnarray}
Thus,
\begin{eqnarray}
\max_{\bar{E}\in\calH(E)}\sum_{v\in\calV}p(v)S_v(E_v)
&\ge&\sum_{v\in\calV}p(v)S_v(E_v^*)\nonumber\\
&=&\sum_{v\in\calV}p(v)[\beta^* E_v^*+\ln Z_v(\beta^*)]\nonumber\\
&=&\beta^*\sum_{v\in\calV}p(v)E_v^*+\sum_vp(v)\ln Z_v(\beta^*)\nonumber\\
&=&\beta^*E+\sum_vp(v)\ln Z_v(\beta^*)\nonumber\\
&=&\bar{S}(E).
\end{eqnarray}
This completes the proof of Theorem 1.
$\Box$

The function $Z_v(\beta)$ is similar to the well--known partition function 
pertaining to the Boltzmann distribution w.r.t.\ the Hamiltonian (energy function) 
$\calE_v(u)=f(u,v)$,
except that each exponential term
is weighted by $q(u|v)$, as opposed to the usual form, 
which is just $\sum_{u\in\calU}e^{-\beta \calE_v(u)}$.
Before describing the statistical mechanical interpretation of eq.\ (\ref{identity}),
we should note that $Z_v(\beta)$ defined in (\ref{zvb}) can easily be related to
the ordinary partition function, without weighting, as follows:
Suppose that $\{q(u|v)\}$ are rational\footnote{Even
if not rational, they can always be approximated as such to an arbitrarily good precision.}
and hence can be represented as ratios
of two positive integers, $q(u|v)=M(u|v)/M$, 
where $M \ge |\calU|$ is common to all $u\in\calU$ (and $v\in\calV$). Now,
imagine that every value of $u$ actually represents
a `quantization' of a more refined microstate (call it a
``nanostate'') $w\in\calW$, $|\calW|=M$, so that $u=g_v(w)$,
where $g_v$ is a many--to--one function, for which the inverse image of every $u$ consists of
$M(u|v)$ many values of $w$. Suppose further that the Hamiltonian depends
on $w$ only via $g_v(w)$, i.e., $\calE_v'(w)=\calE_v(g_v(w))$. Then, the (ordinary) partition
function related to $w$ is given by
\begin{eqnarray}
\label{wpf}
\zeta_v(\beta)&=&\sum_{w\in\calW}e^{-\beta\calE_v'(w)}\nonumber\\
&=&\sum_{w\in\calW}e^{-\beta\calE_v(g_v(w))}\nonumber\\
&=&\sum_{u\in\calU}M(u|v)e^{-\beta\calE_v(u)}\nonumber\\
&=&M\sum_{u\in\calU}q(u|v)e^{-\beta\calE_v(u)}=MZ_v(\beta).
\end{eqnarray}
Thus, the weighted partition function is, within a constant factor $M$,
the same as the ordinary partition function of $w$. This factor
cancels out when probabilities are calculated since it appears both in the
numerator and the denominator. Moreover, 
it affects neither the minimizing $\beta$
that achieves $S_v(E_v)$ or 
$\bar{S}(E)$, nor the derivatives of the log--partition
function.

We now move on to our interpretation of
eq.\ (\ref{identity}) from the viewpoint
of elementary statistical physics: Consider a physical system which consists of $|\calV|$ subsystems of
particles. The total number of particles in the system is $n$ and the total amount
of energy is $nE$ Joules. For each $v\in\calV$, the subsystem indexed by $v$ 
(subsystem $v$, for short) contains
$n_v=np(v)$ particles, each of which can lie in any microstate 
within a finite set of microstates $\calU$ (or an underlying
nanostate in a set $\calW$),
and it is characterized by an additive Hamiltonian 
$\calE_v(u_1,\ldots,u_{n_v})=\sum_{i=1}^{n_v}f(u_i,v)$. The total amount of
energy possessed by subsystem $v$ is given by $n_vE_v$ Joules. As long
as the subsystems are in thermal isolation from each other, each one of them
may have its own temperature $T_v=1/(k\beta_v)$, where $\beta_v$ is the achiever
of the normalized (per--particle) entropy 
associated with an average per--particle energy $E_v$, i.e.,
$$S_v(E_v)=\min_{\beta\ge 0}[\beta E_v+\ln Z_v(\beta)].$$
The above--mentioned rate function $I(E)$ of $\mbox{Pr}\{\sum_{i=1}^nf(U_i,v_i)\le nE\}$
is then given by the negative maximum total per--particle entropy,
$\sum_vp(v)S_v(E_v)$, where the maximum is over all energy allocations $\{E_v\}$
such that the total energy is conserved, i.e., $\sum_vp(v)E_v=E$. 
This maximum is attained by the expression of the
r.h.s.\ of eq.\ (\ref{identity}), where there 
is {\it only one} temperature parameter, and hence
it corresponds to {\it thermal equilibrium}.
In other words, the whole system then lies in the same 
temperature $T^*=1/(k\beta^*)$, where $\beta^*$
is the minimizer of 
$\bar{S}(E)$. Thus, the energy allocation among
the various subsystems in 
equilibrium is such that their temperatures are the same
(cf.\ the above proof of Theorem 1). Theorem 1 is then 
interpreted as expressing the second law
of thermodynamics.

At this point, a few comments are in order:
\begin{enumerate}
\item It should be pointed out that in the above physical interpretation, we have implicitly assumed that the 
particles within each subsystem are distinguishable, and so the partition function corresponding to a set of $n_v$
particles is given by the partition function of a single particle raised to the power of $n_v$, without dividing
by $n_v!$. This differs then from the indistinguishable case only by a constant factor 
(as long as $n_v$ is indeed constant)
and hence the difference between the distinguishable and the indistinguishable 
cases is not essential for the most part of our discussion.
\item As mentioned in the above paragraph, our conclusion is that $I(E)=-\bar{S}(E)$. At first glance, this may
seem peculiar as it appears that $I(E)$ may be negative. However, one should keep in mind that $\bar{S}(E)$
is induced by a (convex) combination of weighted partition functions, 
rather than ordinary partition functions, like $\zeta_v(\beta)$. Referring to eq.\ (\ref{wpf}), the ordinary
notion of entropy $\Sigma(E)$ 
as the normalized log--number of (nano)states with normalized energy $E$, is
given by
\begin{eqnarray}
\bar{\Sigma}(E)&=&\min_{\beta\ge0}\left[\beta E+
\sum_vp(v)\ln \zeta_v(\beta)\right]\nonumber\\
&=&\min_{\beta\ge 0}\left[\beta E+\sum_vp(v)\ln Z_v(\beta)\right]+\ln M\nonumber\\
&=&\bar{S}(E)+\ln M.
\end{eqnarray}
Thus,
$$I(E)=\ln M - \bar{\Sigma}(E),$$
which is always non--negative.
\item The identity (\ref{identity})
can be thought
of as a generalized concavity property of the entropy: 
Had all the entropy 
functions $S_v(\cdot)$ been the same, this would have been the usual
concavity property. What makes this equality less trivial and more interesting
is that it continues to hold even when $S_v(\cdot)$, for 
the various $v\in\calV$, are different from each
other.
\item On the more technical level, since this paper draws analogies with physics,
we should say a few words about physical units. The products $\beta E$, $\beta E_v$, $\beta f(u,v)$, etc.,
should all be pure numbers, of course. Since $\beta=1/(kT)$,
where $k$ is Boltzmann's constant and $T$ is absolute temperature,
and since $kT$ has units of energy (Joules or ergs, etc.),
it is understood that $E$, $E_v$, $f(u,v)$ and the like, should all have units of energy as well. In the applications
described below, whenever this is not the case, i.e., the latter quantities are pure numbers rather than physical energies,
we will sometimes reparametrize $\beta$ by $\beta\epsilon_0$, where $\epsilon_0$ is an arbitrary constant possessing
units of energy (e.g., $\epsilon_0=1$ Joule or $\epsilon_0=1$ erg), 
and we absorb $\epsilon_0$ in the Hamiltonian, i.e., 
redefine $\calE_v(u)=\epsilon_0f(u,v)$. Thus, in this case, $S_v(E)$, where $E$ is the now the energy in units 
of $\epsilon_0$, is redefined as
$$S_v(E)=\min_{\beta\ge 0}
\left[\beta\cdot\epsilon_0 E+
\ln\left(\sum_uq(u|v)e^{-\beta\calE_v(u)}\right)\right].$$
This kind of modification is
not essential, but it may help to avoid confusion 
about units when the picture
is viewed from the aspects of physics.
\end{enumerate}

\section{Applications}

Equipped with the main result of the previous section and its statistical mechanical
interpretation, we next introduce a few applications that fall within the 
framework considered. In all these applications,
there is an underlying large deviations event of the type of eq. (\ref{ld}), whose rate function
is of interest. The above described viewpoint of statistical physics is then relevant in all
these applications.

\subsection{The Rate--Distortion Function}

Let $P=\{p(x),~x\in\calX\}$ designate the vector 
of letter probabilities associated with a
given discrete memoryless source (DMS), and for a given reproduction 
alphabet $\hat{\calX}$, let $d:\calX\times\hat{\calX}\to\reals^+$
denote a single--letter distortion measure. Let $R(D)$ denote the rate--distortion
function of the DMS $P$.

One useful way to think of the rate--distortion function
is inspired by the classical random coding argument:
Let $(\hX_1,\ldots,\hX_n)$ be drawn i.i.d.\ from the
optimum random coding distribution $q^*(\hx_1,\ldots,\hx_n)=\prod_{i=1}^n
q^*(\hx_i)$ and 
consider the event $\sum_{i=1}^n d(x_i,\hX_i)\le nD$,
where $x^n$ is a given source vector, typical to $P$, i.e.,
the composition of $x^n$ consists 
of $n_x=np(x)$ occurrences of each $x\in\calX$. This is exactly an event of the type (\ref{ld}) with
$U_i=\hX_i$, $v_i=x_i$, $i=1,\ldots,n$, $q(u|v)=q(\hx|x)=q^*(\hx)$ independently of $x$,
$f(u,v)=f(\hx,x)=d(x,\hx)$, and $E=D$. I.e., the Hamiltonian $\calE_x(\hx)$ is given by
$\epsilon_0d(x,\hx)$ and the total energy is $nD$ in units of $\epsilon_0$.

Suppose that this
probability is of the exponential order of $e^{-nI(D)}$. Then,
it takes about $M=e^{n[I(D)+\epsilon]}$ ($\epsilon > 0$, however small)
independent trials to `succeed' at least once
(with high probability) in having some realization of $\hX^n$ 
within distance $nD$ from $x^n$. 
This is the well--known
the classical random coding achievability argument that leads to $I(D)=R(D)$. 
Thus, the large--deviations rate function of interest agrees exactly 
with the rate--distortion function (cf.\ \cite[Sect.\ 3.4]{Berger71}), which is:
\begin{equation}
R(D)=-\min_{\beta\ge 0}\left[\beta\cdot\epsilon_0D+\sum_{x\in\calX}p(x)
\ln\left(\sum_{\hx\in\hat{\calX}}q^*(\hx)e^{-\beta\cdot\epsilon_0 d(x,\hx)}\right)\right].
\end{equation}
Interestingly,
in \cite[p.\ 90, Corollary 4.2.3]{Gray90}), the rate--distortion function
is shown, using completely different considerations,
to have a parametric representation which can be written exactly in this form.

The fact that the rate--distortion function has an 
interpretation of an isothermal equilibrium situation in
statistical thermodynamics is not quite new 
(cf.\ e.g.\ \cite[Sect.\ 6.4]{Berger71}, \cite{Rose94}).
Here, however, we obtain it in a more explicit 
manner and as a special case of a more
general principle.

A simple example is that of the binary symmetric source with the Hamming distortion
measure. It is easy to see that, in this example, 
the relationship between distortion and temperature is:
\begin{equation}
T=\frac{\epsilon_0}{k\ln[(1-D)/D]}~~\mbox{or, equivalently,}~~D=\frac{1}{1+e^{\epsilon_0/(kT)}}
\end{equation}
and, of course, $R(D)=1-h_2(D)$, where $h_2(D)$ is the binary entropy function.

A slightly more involved example pertains to the regime
of high resolution (small distortion) and it turns out to 
be related to (a generalized version of) the 
law of equipartition of energy in statistical physics:
Consider the $L_\theta$ distortion measure, $d(x,\hx)=|x-\hx|^\theta$ (most
commonly encountered are the cases $\theta=1$ and $\theta=2$). Let
us assume that $D > 0$ is very small and consider the (continuous)
uniform random coding distribution $q(\hx)=\frac{1}{2A}$ in the interval
$[-A,A]$ and zero elsewhere. This random coding distribution is suboptimal, but
it corresponds, and hence is well motivated,
by many results in high--resolution quantization using
uniform quantizers (see, e.g., \cite{GN98} and references therein).
For every $x\in\calX$, the partition function
is given by
$$Z_x(\beta)=\frac{1}{2A}\int_{-A}^A
\exp\{-\beta\epsilon_0|\hx-x|^\theta\} \mbox{d}\hx.$$
When $D$ is very small, $\beta$ is very large, and
then the finite--interval integral pertaining to $Z_x(\beta)$ can be
approximated\footnote{See the Appendix (subsection A.2) 
for a more rigorous derivation.} by an infinite one,
provided that the support of $\{p(x)\}$ 
is included\footnote{An alternative, softer
condition is that the probability that $|x|\ge A$ is negligibly small.}
in the interval $[-A,A]$:
\begin{equation}
\label{approxz}
Z_x(\beta)\approx\frac{1}{2A}\int_{-\infty}^\infty
\exp\{-\beta\epsilon_0|\hx-x|^\theta\} \mbox{d}\hx,
\end{equation}
which then becomes independent of $x$. The average distortion
(internal energy) associated with this partition function can
be evaluated using the same technique as the one that leads to the
law of equipartition in statistical physics:
\begin{eqnarray}
\label{equipartition}
\epsilon_0 D&\approx&-\frac{\partial}{\partial \beta}\ln
\left[\int_{-\infty}^\infty
\exp\{-\beta\epsilon_0|\hx-x|^\theta\} \mbox{d}\hx\right]\nonumber\\
&=&-\frac{\partial}{\partial \beta}\ln\left[
\beta^{-1/\theta}\cdot\int_{-\infty}^\infty
\exp\{-\epsilon_0|\beta^{1/\theta}(\hx-x)|^\theta\}
\mbox{d}(\beta^{1/\theta}(\hx-x))\right]\nonumber\\
&=&-\frac{\partial}{\partial \beta}\ln\left[
\beta^{-1/\theta}\cdot\int_{-\infty}^\infty
\exp\{-\epsilon_0|z|^\theta\}
\mbox{d}z\right]\nonumber\\
&=&-\frac{\mbox{d}}{\mbox{d}\beta}\ln
\left(\beta^{-1/\theta}\right)-
\frac{\partial}{\partial \beta}\ln\left[
\int_{-\infty}^\infty
\exp\{-\epsilon_0|z|^\theta\}
\mbox{d}z\right]\nonumber\\
&=&\frac{1}{\beta \theta}-0
=\frac{kT}{\theta}
\end{eqnarray}
[Note that for $\theta=2$, where the Hamiltonian is 
quadratic in the integration variable 
$\hx$, this is exactly the law of equipartition.]
Thus, for low temperatures, the distortion
is given by $D=kT/(\epsilon_0\theta)$, i.e.,
distortion is linear in temperature in that regime,
and the constant of proportionality is related to the 
heat capacity, $C=k/\theta$. 
Since the temperature is proportional to the negative local slope
of the distortion--rate function (as the reciprocal, $\beta$, is proportional
to the negative local slope of the rate--distortion function), this means that the distortion
is proportional to its derivative w.r.t.\ $R$, which means an exponential relationship of the
form $D=D_0e^{-\theta R}$ ($D_0$ -- constant). For $\theta=2$ (mean square error), 
this is recognized as the well--known characterization
of distortion as function of rate in the high resolution regime.
Specifically, in this case, the factor of $2$ at the denominator
of $kT/2$, the universal expression of the 
internal energy per degree of freedom according to
the equipartition theorem, has the same origin as the factor of $2$ that appears 
in the exponent
of $D(R)=D_0e^{-2R}$ (decay of 6dB per bit).
Thus the law of equipartition in statistical physics is 
related to the behavior of rate distortion codes in the high resolution regime.

To compute the rate associated with this temperature more explicitly, 
note that the minimizing $\beta^*$ 
is given by $1/(\theta\epsilon_0D)$, and so
\begin{eqnarray}
R&=&-\beta^*\epsilon_0D-
\ln\left[\frac{1}{2A}\int_{-\infty}^\infty
\exp\{-\beta^*\epsilon_0|\hx-x|^\theta\} \mbox{d}\hx\right]\nonumber\\
&=&-\frac{1}{\theta}-\ln\left[\frac{1}{2A}
\cdot\frac{2\Gamma(1/\theta)}{\theta(1/\theta D)^{1/\theta}}\right]\nonumber\\
&=&\ln\left[\frac{A\theta}{\Gamma(1/\theta)
(\theta eD)^{1/\theta}}\right]\nonumber\\
&=&\ln\left[\frac{A\theta}{\Gamma(1/\theta)}\right]-
\frac{1}{\theta}\ln(\theta eD).
\end{eqnarray}

\subsection{Channel Capacity}

In complete duality to the random coding argument that puts the
rate--distortion function in the framework discussed in Section 3,
a parallel argument can be made with regard to channel capacity.

Given a discrete memoryless channel (DMC) with a finite input alphabet $\calX$, 
and a finite output alphabet $\calY$, we can obtain capacity using the following
argument. Let $\{q^*(x),~x\in\calX\}$ be the optimum random coding distribution according
to which, each codeword $X^n$ is drawn independently. Let $y^n$ be a given channel
output sequence which is typical to the output distribution $p(y)=\sum_{x\in\calX}q(x)W(y|x)$,
where $\{W(y|x),~x\in\calX,~y\in\calY\}$ are the channel transition probabilities. That is,
each symbol $y$ appears $n_y=np(y)$ times in $y^n$. Consider now the large deviations event
\begin{equation}
\label{td}
\sum_{i=1}^n\log\frac{1}{W(y_i|X_i)}\le nH(Y|X),
\end{equation}
where $H(Y|X)=-\sum_{x\in\calX}\sum_{y\in\calY}q(x)W(y|x)\log W(y|x)$.
By the union bound, as long as the number of randomly chosen codewords is exponentially less
than $e^{-nI}$, where $I$ is the rate function of the large--deviations event (\ref{td}), then
the average error probability still vanishes as $n\to\infty$.\footnote{Here we apply the union
bound to a threshold decoder that seeks a unique codeword that satisfies (\ref{td}),
which although suboptimum, is still good enough to achieve capacity.} 
Since this is the exactly the achievability argument of the channel coding theorem, then $I=C$, where $C$ the channel
capacity. 

Again, this complies with our model setting with the assignments, $U_i=X_i$, $v_i=y_i$, $i=1,\ldots,n$,
$q(u|v)=q(x|y)=q^*(x)$ independently of $y$,
$f(u,v)=f(x,y)=-\log W(y|x)$ and $E=H(Y|X)$ units of $\epsilon_0$. 
In other words, channel capacity can be represented as
\begin{equation}
C=-\min_{\beta\ge 0}\left[\beta \cdot \epsilon_0 H(Y|X)+
\sum_{y\in\calY}p(y)\ln\left(\sum_{x\in\calX}q^*(x)e^{-\beta\cdot\epsilon_0[-\log W(y|x)]}\right)\right].
\end{equation}
It is easy to see that, in this case, the equilibrium 
temperature always corresponds 
to $\beta\epsilon_0=1$, namely, $T=\epsilon_0/k$.

By the same token, one can derive an expression 
of the random coding capacity pertaining to mismatched
decoding, where the decoder uses an additive metric $m(x,y)$ 
other than the optimum metric, 
$-\log W(y|x)$ (see, e.g., \cite{Balakirsky95}, 
\cite{CN95},
\cite{Lapidoth94},
\cite{LS96-2}, 
\cite{MKLS94}, 
and references therein).
The only modifications to the above 
expression would be to replace the Hamiltonian
by $\calE_y(x)=\epsilon_0m(x,y)$ 
and to replace $H(Y|X)$ by the expectation
of $m(X,Y)$ w.r.t.\ $q^*(x)W(y|x)$. 
The new optimum random coding distribution
might change as well. Here, it
is no longer necessarily true that the equilibrium temperature
is $T=\epsilon_0/k$.

\subsection{Signal Detection and Hypothesis Testing}

Consider the following binary hypothesis testing problem:
Given a deterministic signal, which is repreresented by a sequence $x^n=(x_1,\ldots,x_n)$
with elements taking on values in a (finite) set $\calX$ and relative frequencies $\{p(x),~x\in\calX\}$, 
and given an observation sequence
$Y^n=(Y_1,\ldots,Y_n)$, we are required to decide between two hypotheses:
\begin{itemize}
\item[$H_0:$] The observation vector $Y^n$ is ``pure noise,'' 
distributed according to some product measure $Q=\{q(y),~y\in\calY\}$, i.e., $q(y^n)=\prod_{i=1}^nq(y_i)$,
which is unrelated to $x^n$.
\item[$H_1:$] The observation vector $Y^n$ is a ``noisy version'' of $x^n$, 
distributed according to $q(y^n|x^n)=\prod_{i=1}^nq(y_i|x_i)$.
\end{itemize}
The optimum detector (under both the Bayesian and the Neyman--Pearson criterion) compares
the likelihood ratio $\sum_{i=1}^n\ln [q(y_i)/q(y_i|x_i)]$ to a threshold $nE_0$, and decides
in favor of $H_0$ if this threshold is exceeded, otherwise, it decides in favor of $H_1$.

The false--alarm probability then is the probability of the event
$$\sum_{i=1}^n\ln \left[\frac{q(Y_i)}{q(Y_i|x_i)}\right]\le nE_0$$
under $Q$. This, again, fits our scenario with the substitutions 
$U_i=Y_i$, $v_i=x_i$, $i=1,\ldots,n$, 
$q(u|v)=q(y)$, independently of $x=v$, $f(u,v)=f(y,x)=\ln[q(y)/q(y|x)]$, and $E=E_0$. 
Similarly, the analysis of the missed--detection probability corresponds to
the assignments: $U_i=Y_i$, and $v_i=x_i$, $i=1,\ldots,n$, as before, but now
$q(u|v)=q(y|x)$, $f(u,v)=f(y,x)=\ln[q(y|x)/q(y)]$ and $E=-E_0$.
Note that when $\{q(y)\}$
is the uniform distribution over $\calY$, the missed-detection event
can also be interpreted as the probability of excess code--length of
an arithmetic lossless source code w.r.t.\ $\{q(y|x)\}$.

Another situation of hypothesis testing that is related to our study in a similar manner is 
one where the signal $x^n$ is always underlying the observations, but the decision to be made
is associated with two hypotheses regarding 
the noise level, or the temperature. In this case, there is a certain
Hamiltonian $\calE_x(y)$ for each $x\in\calX$, and we assume a Boltzmann--Gibbs distribution
parametrized by the temperature
$$q(y|x,\beta)=\frac{e^{-\beta\calE_x(y)}}{\zeta_x(\beta)}$$
where
$$\zeta_x(\beta)=\sum_ye^{-\beta\calE_x(y)}.$$
Note that here $\zeta_x(\beta)$ is an ordinary partition function, without
weighting (cf.\ (\ref{wpf})). We shall also denote
$$\bar{\Sigma}(E)=\min_{\beta\ge 0}\left[\beta E
+\sum_{x\in\calX}p(x)\ln\zeta_x(\beta)\right].$$
As $\bar{\Sigma}(E)$ is induced by a convex combination of non-weighted
partition functions, it has the significance of the normalized logarithm
of the number of microstates with energy about $nE$. Thus, $k\cdot\bar{\Sigma}(E)$,
where $k$ is Boltzmann's constant, is the thermodynamic entropy.

Given two values $\beta_1$ and $\beta_2$ (say, $\beta_1 > \beta_2$),
the hypotheses now are the following:
\begin{itemize}
\item[$H_1:$] $Y^n$ is 
distributed according to $q_1(y^n|x^n)=\prod_{i=1}^nq(y_i|x_i,\beta_1)$.
\item[$H_2:$] $Y^n$ is 
distributed according to $q_2(y^n|x^n)=\prod_{i=1}^nq(y_i|x_i,\beta_2)$.
\end{itemize}
The likelihood ratio test compares 
$\sum_{i=1}^n\calE_{x_i}(Y_i)$ to a threshold, $nE_0$, and 
decides in favor of $H_2$ if the threshold 
is exceeded, otherwise, it favors $H_1$. 
Here, $E_0$ should lie in the interval $(E_1,E_2)$,
where
$$E_i\dfn-\sum_{x\in\calX}p(x)\cdot\left[\frac{\partial\ln \zeta_x(\beta)}{\partial
\beta}\right]_{\beta=\beta_i},~~~i=1,2.$$
For convenience, let us assume now that 
$E_i$, $i=0,1,2,$ and $\calE_x(y)$ already have units of energy, so 
there is no need to have the constant $\epsilon_0$. In this
situation, the exponent of the error probability under $H_2$ is given by
$-\bar{S}(E_0)$, where
\begin{eqnarray}
\bar{S}(E_0)&=&\min_{\beta\ge 0}\left[\beta E_0
+\sum_{x\in\calX}p(x)\ln\left(\sum_{y\in\calY}q(y|x,\beta_2)
e^{-\beta\calE_x(y)}\right)\right]\nonumber\\
&=&\min_{\beta\ge 0}\left[\beta E_0+
\sum_{x\in\calX}p(x)\ln\left(\frac{\zeta_x(\beta+\beta_2)}
{\zeta_x(\beta_2)}\right)
\right]\nonumber\\
&=&\min_{\beta\ge 0}\left[\beta E_0+
\sum_{x\in\calX}p(x)\ln \zeta_x(\beta+\beta_2)-
\sum_{x\in\calX}p(x)\ln \zeta_x(\beta_2)
\right]\nonumber\\
&=&\min_{\beta\ge 0}\left[(\beta+\beta_2)E_0+
\sum_{x\in\calX}p(x)\ln \zeta_x(\beta+\beta_2)\right]-\beta_2E_0-
\sum_{x\in\calX}p(x)\ln \zeta_x(\beta_2)\nonumber\\
&=&\min_{\beta\ge \beta_2}\left[\beta E_0+
\sum_{x\in\calX}p(x)\ln \zeta_x(\beta)\right]+\beta_2(E_2-E_0)
-\left[\beta_2E_2+\sum_{x\in\calX}p(x)\ln \zeta_x(\beta_2)\right]\nonumber\\
&=&\min_{\beta\ge \beta_2}\left[\beta E_0+
\sum_{x\in\calX}p(x)\ln \zeta_x(\beta)\right]+\beta_2(E_2-E_0)\nonumber\\
& &-\min_{\beta\ge 0}\left[\beta E_2+\sum_{x\in\calX}p(x)\ln \zeta_x(\beta)\right]\nonumber\\
&=&\bar{\Sigma}(E_0)-\bar{\Sigma}(E_2)+\beta_2(E_2-E_0),
\end{eqnarray}
where we have used the fact that the achiever $\beta(E)$
of $\bar{\Sigma}(E)$ is a monotonically non-increasing function of $E$,
thus, $E_0 < E_2$ implies $\beta(E_0) > \beta(E_2)=\beta_2$,
and so, the global minimum over $\beta\ge 0$ is attained
for $\beta\ge\beta_2$ anyway.

It then follows that the error exponent $I_2$ under $H_2$ is given by
\begin{eqnarray}
I_2&=&\bar{\Sigma}(E_2)-\bar{\Sigma}(E_0)-\beta_2(E_2-E_0)\nonumber\\
&=&\frac{1}{k}\left[k\bar{\Sigma}(E_2)-k\bar{\Sigma}(E_0)-
\frac{E_2-E_0}{T_2}\right]\nonumber\\
&=&\frac{1}{k}\int_{E_0}^{E_2}\left[\frac{1}{T(E)}-\frac{1}{T_2}\right]\mbox{d}E
\nonumber\\
&=&\frac{1}{k}\int_{T_0}^{T_2}\left(\frac{1}{T}-\frac{1}{T_2}\right)
\bar{C}(T)\mbox{d}T,
\end{eqnarray}
where $T(E)=1/(k\beta(E))$ is the temperature corresponding to
energy $E$, $T_i=T(E_i)$, $i=0,1,2$, and $\bar{C}(T)=\mbox{d}E/\mbox{d}T$
is the average heat capacity per particle of the system, which is
the weighted average of heat capacities of all subsystems, i.e.,
$$\bar{C}(T)=\sum_{x\in\calX}p(x)C_x(T),$$
where
$$C_x(T)=\frac{\mbox{d}E_x}{\mbox{d}T}=
\frac{1}{kT^2}\left[\frac{\mbox{d}^2\ln \zeta_x(\beta)}{d\beta^2}\right]_{\beta=
1/(kT)}.$$
Thus,
$$I_2=\sum_{x\in\calX}p(x)\cdot
\frac{1}{k}\int_{T_0}^{T_2}\left(\frac{1}{T}-\frac{1}{T_2}\right)
C_x(T)\mbox{d}T,$$
which is interpreted as the weighted average of the relative contributions
of all subsystems, which all lie in the same temperature $T_0$.

In a similar manner, the rate function $I_1$ of the probability
of error under $H_1$ is given by:
\begin{eqnarray}
I_1&=&\bar{\Sigma}(E_1)-\bar{\Sigma}(E_0)-\beta_1(E_1-E_0)\nonumber\\
&=&\frac{1}{k}\left[
k\bar{\Sigma}(E_1)-k\bar{\Sigma}(E_0)-\frac{E_1-E_0}{T_1}\right]\nonumber\\
&=&\frac{1}{k}\int_{E_1}^{E_0}\left[\frac{1}{T_1}-
\frac{1}{T(E)}\right]\mbox{d}E\nonumber\\
&=&\frac{1}{k}\int_{T_1}^{T_0}\left(\frac{1}{T_1}-
\frac{1}{T}\right)\bar{C}(T)\mbox{d}T.
\end{eqnarray}

The expression in the square brackets 
of the second line pertaining to $I_2$ has a simple graphical 
interpretation (see Fig.\ 1): It is the vertical distance 
(corresponding to the vertical line $E=E_0$) between the curve $\bar{\Sigma}(E)$
and the line tangent to that curve at $E=E_2$ (whose slope is $\beta_2=
\beta(E_2)$). The two other expressions of $I_2$, in the 
last chain of equalities, describe the error exponent $I_2$ in terms
of slow heating from temperature $T_0$ to temperature $T_2$. 
Similar comments apply to $I_1$ (cf.\ Fig.\ 1).
Thus, the error exponents
are linear functionals of the average heat capacity, $\bar{C}(T)$,
in the range of temperatures $[T_1,T_2]$. 
The higher is the heat capacity, the better is the discrimination
between the hypotheses. This is related to the fact that Fisher information
of the parameter $\beta$ is given by
$$J(\beta)=\sum_{x\in\calX}p(x)\frac{\mbox{d}^2\ln \zeta_x(\beta)}{\mbox{d}\beta^2}=
kT^2\bar{C}(T),$$
namely, again, a linear function of $\bar{C}(T)$. 
However, while the Fisher information
depends only on one local value of $\bar{C}(T)$ 
(as it measures the sensitivity of the
likelihood function to the parameter in a local manner), the error exponents
depend on $\{\bar{C}(T): T_1\le T\le T_2\}$ 
in a cumulative manner, via the above integrals.
The tradeoff between $I_1$ and $I_2$ is also obvious: by enlarging the
threshold $E_0$, or, correspondingly, $T_0$, 
the range of integration pertaining to
$I_1$ increases at the expense of the one of $I_2$ and vice versa. In the
extreme case, where $I_2=0$, we get
$$I_1=D(P_2\|P_1)=
\frac{1}{k}\int_{T_1}^{T_2}\left(\frac{1}{T_1}-
\frac{1}{T}\right)\bar{C}(T)\mbox{d}T.$$

\begin{figure}[ht]
\hspace*{1cm}\input{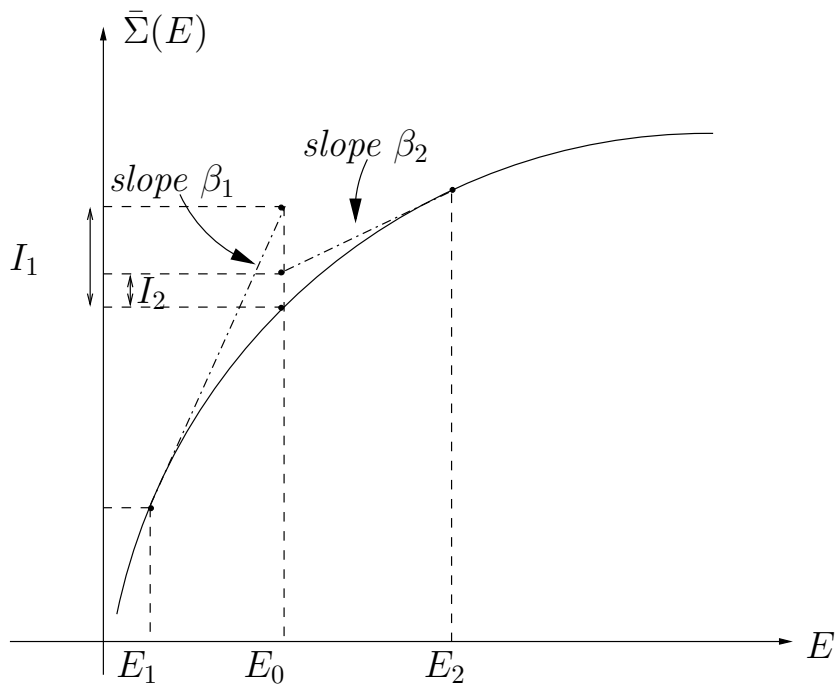}
\caption{Entropy as function of energy and a graphical representation 
of error exponents.}
\label{gen}
\end{figure}

\subsection{Error Exponents of Time--Varying Scalar Quantizers}

In this application example, we are back to the problem area
of lossy data compression, but this time, it is about scalar (symbol--by--symbol)
compression. This setup is motivated by earlier results about the optimality
of time--shared scalar quantizers within the class of 
causal source codes for memoryless sources, both under
the average rate/distortion criteria \cite{NG82} and large--deviations performance
criteria \cite{MK03}. In particular, it was shown that 
under both criteria, optimum time--sharing
between at most two (entropy coded) scalar quantizers
is as good as any causal source code for memoryless sources.
Here, we will focus on the large deviations performance criteria, namely,
source coding exponents.

Consider a time--varying scalar quantizer $\hX_i=f_i(X_i)$, acting on a DMS
$X_1,X_2,\ldots$, $X_i\in\calX$, drawn from $q$, 
where $\{f_i\}$ is an arbitrary (deterministic) sequence
of quantizers from a given finite set $\calF=\{F_1,\ldots,F_S\}$,
where $F_s:\calX\to\hat{\calX}_s$, $\hat{\calX}_s$ being the reproduction alphabet
corresponding to $F_s$, $s=1,\ldots,S$. In other words, for every $i=1,2,\ldots,n$,
$f_i=F_{s_i}$, for a certain arbitrary sequence of `states',
$s_1,s_2,\ldots$ (known to the decoder) with components in $\calS=\{1,2,\ldots,S\}$. 

The distortion incurred by such a time--varying scalar quantizer, over $n$ units of time, is
$\sum_{i=1}^nd(X_i,f_i(X_i))=
\sum_{i=1}^nd(X_i,F_{s_i}(X_i))$. The total code length is $\sum_{i=1}^n L_{s_i}(F_{s_i}(X_i))$,
where the per--symbol length functions
$L_{s}(\cdot)$ may correspond to either fixed--rate coding, where $L_s(\hx)=R_s\dfn
\lceil\log|\hat{\calX}_s|\rceil$ for all $\hx$, 
or any other length function satisfying the Kraft
inequality, $\sum_{\hx\in\hat{\calX}_s}2^{-L_s(\hx)}\le 1$.
For the sake of simplicity of the exposition, let us assume fixed--rate coding.
We will denote by $n_s$, $s\in\calS$, the number of times that $s_i=s$
occurs in $s^n$, and $p(s)=n_s/n$ is the corresponding relative frequency.

In \cite{MK03}, among other results, the rate function of the excess distortion event
$$\sum_{i=1}^nd(X_i,F_{s_i}(X_i)) > nD,~~~~
D> \sum_{(x,s)\in\calX\times\calS}q(x)p(s)d(x,F_s(x))$$
was optimized across the class of all time--varying scalar quantizers (each one
corresponding to a different sequence $s_1,\ldots,s_n$) subject to a code--length 
constraint $\sum_{i=1}^nR_{s_i}\le nR$, or equivalently, $\sum_{s\in\calS}n_sR_s\le nR$,
for a given pair $(D,R)$.

In the notation of our generic model, here we have $U_i=X_i$, $v_i=s_i$, $i=1,\ldots,n$,
$q(u|v)=q(x|s)=q(x)$ independently of $s$, 
$f(u,v)=f(x,s)=-d(x,F_s(x))$, and $E=-D$.\footnote{One
may prefer to redefine 
$f(x,s)=D_{\max}-d(x,F_s(x))$ and $E=D_{\max}-D$, 
where $D_{max}\dfn\max_{x,s}d(x,F_s(x))$, in order to
work with non--negative quantities.} and the excess distortion exponent is of the same form
as before (see also \cite{MK03}). 
Here, however, unlike the previous application examples, we have a degree
of freedom to select the relative frequency of usage, $p(s)$, of each member of $\calF$,
i.e., the time--sharing protocol, but we also have the constraint $\sum_sp(s)R_s\le R$.

From the statistical physics point of view, these additional ingredients mean that
we have a freedom to select the number of particles in each subsystem
(though the total number, $n$, is still fixed), and the additional
constraint, $\sum_sp(s)R_s\le R$, which is actually equivalent to the equality constraint
$\sum_sp(s)R_s= R$ (in the interesting region of $(R,D)$ pairs) can be viewed as an additional
conservation law with respect to some other 
constant of motion, in addition to the energy (e.g., the momentum), where in
subsystem $s$, the (average) value of the corresponding physical quantity 
per particle is $R_s$.

While in \cite{MK03}, we have considered the problem of maximizing the rate function
(the source coding exponent) of the excess distortion event
$\sum_{i=1}^nd(X_i,F_{s_i}(X_i)) > nD$, a related objective (although somewhat 
less well motivated, but still interesting) is to minimize the rate function 
(or maximize the probability) of the small distortion event
$$\sum_{i=1}^nd(X_i,F_{s_i}(X_i)) < nD,~~~
D < \sum_{(x,s)\in\calX\times\calS}q(x)p(s)d(x,F_s(x)).$$
In this case, the optimum performance is given by
$$F(R,D)=\max_{P\in \calP(R)}\min_{\beta\ge 0}\left[\beta D+\sum_{s=1}^Sp(s)\ln
\left(\sum_{x\in\calX}q(x)e^{-\beta d(x,F_s(x))}\right)\right],$$
where $\calP(R)$ is the class of all probability distributions $P=\{p(s),~s\in\calS\}$ with
$\sum_sp(s)R_s\le R$. From the viewpoint of statistical physics, this corresponds
to a situation where the various subsystems are allowed to interact, not only thermally,
but also chemically, i.e., an exchange of particles is enabled in addition to the exchange of
energy, and the maximization over $\calP(R)$ (maximum entropy) is achieved when the
chemical potentials of the various subsystems reach a balance. As the maximization over
$P\in\calP(R)$ subject to the constraint $\sum_sp(s)R_s\le R$, for a given $\beta$, 
is a linear programming
problem with one constraint (in addition to $\sum_sp(s)=1$), then as was shown in
\cite{MK03}, for each distortion level (or energy) $D$, the optimum $P\in\calP(R)$ may be
non--zero for at most two members of $\calS$ only, which means that at most two subsystems
are populated by particles in thermal and chemical equilibrium under the two conservation
laws (of $D$ and of $R$). However, the choice of these two
members of $\calS$ depends, in general, 
on $D$, which in turn depends on the temperature. Thus, when
the system is heated gradually, certain {\it phase transitions} 
may occur, whenever
there is a change in the choice of the two populated subsystems.

Finally, referring to comment no.\ 1 of Section 3, we should point out that here,
in contrast to our discussion thus far, the difference between the ensemble of
distinguishable particles and indistinguishable particles becomes critical since the
factors $\{n_s!\}$ are no longer constant. Had we assumed indistinguishability, the
normalized log--partition function 
would no longer be affine in $P$, thus the maximization over $P$
would no longer be a linear programming problem, and the conclusion might have been
different. In the source coding problem, the indistinguishable case corresponds to
a situation where the sequence of states $s^n$ is chosen uniformly at random
(with the decoder being informed of the result 
of the random selection, of course). In this case,
the Chernoff bound corresponding to each composition $\{n_s,~s\in\calS\}$ of $s^n$
should be weighed by the probability of this composition, which is
$S^{-n}n!/\prod_sn_s!$. Now, each factor of $1/n_s!$ can be
absorbed in the corresponding 
partition function $Z_s(\beta)$ of subsystem $s$, with the interpretation
that in each subsystem the particles are now indistinguishable. The maximum over $P$ would
now correspond to the dominant contribution in this weighted average of Chernoff bounds.
One can, of course, extend the discussion to any i.i.d.\ distribution on $s^n$, thus
introducing additional bias and preferring some compositions over others.

\section*{Appendix}
\renewcommand{\theequation}{A.\arabic{equation}}
    \setcounter{equation}{0}

\subsection*{A.1. Sketch of an Alternative Proof of Theorem 1 via Chernoff Bounds}

In this subsection, 
we outline another proof of Theorem 1 
using a large deviations analysis approach. In particular,
consider the large deviations event $\sum_{i=1}^nf(U_i,v_i)\le nE$,
as described in Section 2.
Assuming that the relative frequencies $\{p(v)\}$ all stabilize
as $n\to\infty$, let us compute the rate function $I(E)$
of the probability of this event in two different methods, where one would yield
the left--hand side of (\ref{identity}) and the other would give the right--hand
side of (\ref{identity}).

In the first method,
we partition the sequence $v^n$ according to its different letters.
Specifically, let 
$$E_v\dfn\frac{1}{n_v}\sum_{i:v_i=v}f(U_i,v),$$ 
where $n_v$ is the number of occurrences of the symbol $v\in\calV$ along $v^n$.
Let $\calG$ denote the set of all possible vector values that
can be taken on by the vector $\bar{E}=\{E_v,~v\in\calV\}$.
Now, obviously, $\sum_{i=1}^n f(U_i,v_i)\le nE$
if and only if there exists a 
vector $\tilde{E}=\{\tilde{E}_v,~v\in\calV\}\in\calG$ 
such that $E_v\le \tilde{E}_v$ for all $v\in\calV$ and
$\sum_{v\in\calV}p(v)\tilde{E}_v\le E$. 
The ``if'' part follows from
$$\sum_{i=1}^n f(U_i,v_i)=n\sum_{v\in\calV}p(v)E_v\le
n\sum_{v\in\calV}p(v)\tilde{E}_v\le nE.$$
The ``only if'' part follows by setting $\tilde{E}_v=E_v$ for all $v\in\calV$.
Therefore, denoting 
$\calH_G(E)=\calH_0(E)\bigcap\calG$ (where $\calH_0(E)$ is defined as in Section 2), we have:
\begin{eqnarray}
\label{ub1}
\mbox{Pr}\left\{\sum_{i=1}^n f(U_i,v_i)\le nE\right\}&=&
\mbox{Pr}\bigcup_{\bar{E}\in\calH_G(E)}\left\{\sum_{i:v_i=v} 
f(U_i,v)\le n_v\tilde{E}_v,~~v\in\calV\right\}\nonumber\\
&\le&\sum_{\tilde{E}\in\calH_G(E)}\mbox{Pr}\left\{\sum_{i:v_i=v} 
f(U_i,v)\le n_v\tilde{E}_v,~~v\in\calV\right\}\nonumber\\
&=&\sum_{\tilde{E}\in\calH_G(E)}\prod_{v\in\calV}
\mbox{Pr}\left\{\sum_{i:v_i=v} f(U_i,v)\le n_v\tilde{E}_v\right\}\nonumber\\
&\le&|\calH_G(E)|\cdot\max_{\tilde{E}\in\calH_G(E)}\prod_{v\in\calV}
\mbox{Pr}\left\{\sum_{i:v_i=v} f(U_i,v)\le n_v\tilde{E}_v\right\}\nonumber\\
&\le&|\calG|\cdot\max_{\tilde{E}\in\calH_G(E)}\prod_{v\in\calV}
\mbox{Pr}\left\{\sum_{i:v_i=v} f(U_i,v)\le n_v\tilde{E}_v\right\},
\end{eqnarray}
and on the other hand,
\begin{eqnarray}
\label{lb1}
\mbox{Pr}\left\{\sum_{i=1}^n f(U_i,v_i)\le nE\right\}&=&
\mbox{Pr}\bigcup_{\tilde{E}\in\calH_G(E)}\left\{\sum_{i:v_i=v} 
f(U_i,v)\le n_v\tilde{E}_v,~~v\in\calV\right\}\nonumber\\
&\ge&\max_{\tilde{E}\in\calH_G(E)}\mbox{Pr}\left\{\sum_{i:v_i=v}
f(U_i,v)\le n_v\tilde{E}_v,~~v\in\calV\right\}\nonumber\\
&=&\max_{\tilde{E}\in\calH_G(E)}\prod_{v\in\calV}
\mbox{Pr}\left\{\sum_{i:v_i=v} f(U_i,v)\le n_v\tilde{E}_v\right\}.
\end{eqnarray}
At this point, the only gap between the upper bound (\ref{ub1}) and
the lower bound (\ref{lb1}) is the factor $|\calG|$. The number of different
values that $\tilde{E}_v$ can take does not exceed the number of different
type classes of sequences of length $n_v$ over the alphabet $\calU$,
which is upper bounded by $(n_v+1)^{|\calU|-1}$.
Thus,
\begin{eqnarray}
|\calG|&\le&\prod_{v\in\calV}[n_v+1]^{|\calU|-1}\nonumber\\
&=&\exp\left\{(|\calU|-1)\sum_v\log(n_v+1)\right\}\nonumber\\
&=&\exp\left\{|\calV|\cdot(|\calU|-1)\sum_v\frac{1}{|\calV|}
\log(n_v+1)\right\}\nonumber\\
&\le&\exp\left\{|\calV|\cdot(|\calU|-1)\log\left(\sum_v\frac{1}{|\calV|}
[n_v+1]\right)\right\}\nonumber\\
&=&\exp\left\{|\calV|\cdot(|\calU|-1)\log\left(
\frac{n}{|\calV|}+1\right)\right\}\nonumber\\
&=&\left(\frac{n}{|\calV|}+1\right)^{|\calV|\cdot(|\calU|-1)},
\end{eqnarray}
and therefore $|\calG|$ is only polynomial in $n$, and hence does not affect the
exponential behavior. Now, each one of the terms
$\mbox{Pr}
\{\sum_{i:v_i=v} f(U_i,v)\le n_v\tilde{E}_v\}$ is bounded
exponentially tightly by an individual Chernoff bound, 
$$\exp\left\{n_v\min_{\beta\ge 0}\left[\beta 
\tilde{E}_v+\ln\left(\sum_uq(u|v)e^{-\beta f(u,v)}
\right)\right]\right\},$$
and so, the dominant term of their product is of the exponential order of
$$\max_{\tilde{E}\in\calH_G(E)}\sum_vp(v)\cdot
\min_{\beta\ge 0}\left[\beta \tilde{E}_v+\ln\left(\sum_uq(u)e^{-\beta f(u,v)}
\right)\right]=\max_{\tilde{E}\in\calH_G(E)}\sum_vp(v)S_v(E_v).$$
Finally, as $n_v\to\infty$, the set $\calH_G(E)$ becomes dense in the continuous set $\calH_0(E)$,
and by simple continuity arguments, the maximum over $\calH_G(E)$ tends to the maximum over $\calH_0(E)$.

The other method to evaluate the rate function $I(E)$
is as follows. Let $\ell$ be a fixed positive integer that divides $n$, 
and denote $\ell_v=\ell p(v)$, $v\in\calV$ (assume that $\ell$ is chosen large enough that
$\ell p(v)$ is well approximated by the closest integer with a very small relative error).
Now, re--order the pairs $\{(U_i,v_i)\}$
(periodically), according to the following rule: 
Assuming, without loss of generality, that
$\calV=\{1,2,\ldots,|\calV|\}$, the first 
$\ell_1=\ell p(1)$ symbol pairs of each $\ell$--block of $(u^n,v^n)$
are such that $v=1$, the next $\ell_2=\ell p(2)$ symbol pairs 
of each $\ell$--block are such that
$v=2$, and so on. In other words, 
each $\ell$--block, $v_{(i-1)\ell+1}^{i\ell}=(v_{(i-1)\ell+1},
v_{(i-1)\ell+2},\ldots,v_{i\ell})$, $i=1,2,\ldots,n/\ell$,
consists of the same relative frequencies $\{p(v)\}$
as the entire sequence, $v^n$. Now, for 
the re--ordered sequence of pairs, let us define 
$X_i=\sum_{t=(i-1)\ell+1}^{i\ell}f(U_t,v_t)$,
$i=1,2,\ldots,n/\ell$. 
Obviously, $X_1,X_2,\ldots,X_{n/\ell}$ are i.i.d.\ and therefore
the probability of the large deviations event $\{\sum_{i=1}^{n/\ell}X_i \le \frac{n}{\ell}\cdot
\ell E\}$ can be assessed exponentially tightly by the Chernoff bound
as follows:
\begin{eqnarray}
&&\exp\left\{\frac{n}{\ell}\cdot\min_{\beta\ge 0}\left[\beta\cdot\ell E+\ln\left(
\sum_{u^\ell\in\calU^\ell}q(u^\ell|v^\ell)\exp\left\{-\beta
\sum_{i=1}^{\ell}f(u_i,v_i)\right\}\right)\right]
\right\}\nonumber\\
&=&\exp\left\{\frac{n}{\ell}\cdot\min_{\beta\ge 0}
\left[\beta\cdot\ell E+\ln\left(\prod_{v\in\calV}
\sum_{u^{\ell_v}}q(u^{\ell_v}|v^{\ell_v})
\exp\left\{-\beta\sum_{i=1}^{\ell_v}f(u_i,v)\right\}\right)\right]
\right\}\nonumber\\
&=&\exp\left\{\frac{n}{\ell}\cdot\min_{\beta\ge 0}
\left[\beta\cdot\ell E+\ln\left(\prod_{v\in\calV}
\left[\sum_{u\in\calU}q(u|v)e^{-\beta f(u,v)}\right]^{\ell_v}\right)\right]
\right\}\nonumber\\
&=&\exp\left\{\frac{n}{\ell}\cdot\min_{\beta\ge 0}
\left[\beta\cdot\ell E+\ell\cdot\sum_{v\in\calV}
p(v)\ln\left(\sum_{u\in\calU}q(u|v)e^{-\beta f(u,v)}\right)\right]
\right\}\nonumber\\
&=&\exp\left\{n\cdot\min_{\beta\ge 0}\left[\beta E+\sum_{v\in\calV}
p(v)\ln\left(\sum_{u\in\calU}q(u|v)e^{-\beta f(u,v)}\right)\right]
\right\}\nonumber\\
&=&e^{n\bar{S}(E)}.
\end{eqnarray}
Since both approaches yield exponentially tight 
evaluations of $I(E)$, they must be equal. 

\subsection*{A.2. A More Rigorous Derivation of Eq.\ (\ref{equipartition})}

The exact derivation of eq.\ (\ref{equipartition}) for the finite
interval integration, is as follows:
\begin{eqnarray}
\epsilon_0 D&=-&\frac{\partial}{\partial \beta}\ln
\left[\int_{-A}^A
\exp\{-\beta\epsilon_0|\hx-x|^\theta\} \mbox{d}\hx\right]\nonumber\\
&=&-\frac{\partial}{\partial \beta}\ln\left[
\beta^{-1/\theta}\cdot\int_{-\beta^{1/\theta}(A+x)}^{\beta^{1/\theta}(A-x)}
\exp\{-\epsilon_0|\beta^{1/\theta}(\hx-x)|^\theta\}
\mbox{d}(\beta^{1/\theta}(\hx-x))\right]\nonumber\\
&=&-\frac{\partial}{\partial \beta}\ln\left[
\beta^{-1/\theta}\cdot\int_{-\beta^{1/\theta}(A+x)}^{\beta^{1/\theta}(A-x)}
\exp\{-\epsilon_0|z|^\theta\}
\mbox{d}z\right]\nonumber\\
&=&-\frac{\partial}{\partial \beta}\ln
\left(\beta^{-1/\theta}\right)-\frac{\partial}{\partial \beta}\ln
\left[\int_{-\beta^{1/\theta}(A+x)}^{\beta^{1/\theta}(A-x)}
\exp\{-\epsilon_0|z|^\theta\}\mbox{d}z\right]\nonumber\\
&=&\frac{1}{\beta\theta}\left\{1-\frac{\beta^{1/\theta}
[(A-x)\exp\{-\beta\epsilon_0|A-x|^\theta\}
+(A+x)\exp\{-\beta\epsilon_0|A+x|^\theta\}]}
{\int_{-\beta^{1/\theta}(A+x)}^{\beta^{1/\theta}(A-x)}
\exp\{-\epsilon_0|z|^\theta\}\mbox{d}z}\right\}.
\end{eqnarray}
When $\beta$ is very large, the denominator 
of the second term of the expression 
in the curly brackets of the right--most side, goes to
$\int_{-\infty}^{\infty}
\exp\{-\epsilon_0|z|^\theta\}\mbox{d}z$, which is a constant. 
Now if, in addition, $|x|<A$, then the numerator
tends to zero as $\beta$ grows without bound. 
Thus, the dominant term, for low temperatures, is $1/(\beta\theta)=kT/\theta$.

An exact closed--form expression, for every finite $\beta$, can be derived
for the case $\theta=1$, since in this case, the integral at the denominator has a simple
expression. For example, setting $\theta=1$, and $x=0$ in the above expression, yields:
\begin{eqnarray}
D&=&\frac{1}{\beta\epsilon_0}-\frac{A}{e^{\beta\epsilon_0A}-1}\nonumber\\
&=&\frac{kT}{\epsilon_0}-\frac{A}{e^{\epsilon_0A/(kT)}-1}.
\end{eqnarray}
Note that this expression is valid only in the range where it is monotonically
increasing in $T$. (Beyond this point, the minimizing $\beta$ is no longer the
point of zero derivative).

\end{document}